\documentclass[twocolumn,reprint,amsmath,amssymb, aps,prl]{revtex4-2}
\usepackage[T1]{fontenc}
\usepackage[latin9]{inputenc}
\usepackage{color}
\usepackage{float}

\makeatletter

\usepackage{amsfonts,amssymb}
\usepackage{latexsym}
\usepackage{epsfig}
\usepackage{graphicx}
\usepackage{tikz}
\definecolor{greatblue}{RGB}{40,120,181}
\definecolor{greatred}{RGB}{200,36,35}
\usepackage[colorlinks,linkcolor=greatblue,anchorcolor=blue,citecolor=greatred]{hyperref}

\usepackage{graphicx}
\usepackage{dcolumn}
\usepackage{bm}

\makeatother

\usepackage[english]{babel}

\begin{document}
\preprint{preprintnumbers}{CTU-SCU/2024005}

\title{An alternative to purification in CFT}

\author{Xin Jiang}  
\email{domoki@stu.scu.edu.cn}
\affiliation{College of Physics, Sichuan University, Chengdu, 610065, China}

\author{Peng Wang}
\email{pengw@scu.edu.cn}
\affiliation{College of Physics, Sichuan University, Chengdu, 610065, China}

\author{Houwen Wu}
\email{iverwu@scu.edu.cn}   
\affiliation{College of Physics, Sichuan University, Chengdu, 610065, China}

\author{Haitang Yang}
\email{hyanga@scu.edu.cn}
\affiliation{College of Physics, Sichuan University, Chengdu, 610065, China}

\date{\today}

\begin{abstract}
In conformal field theories, in contrast to \emph{adding} some auxiliary states into the bipartite  mixed state $\rho_{AB}$  as the
usual purifications do, we show a pure entangled state $\psi_{AB}$ can be constructed by \emph{subtracting} the  undetectable regions. In this pure state $\psi_{AB}$, the von Neumann entropy $S_{\text{vN}}(A)$ naturally captures   quantum entanglement between $A$ and $B$. We verify that  $S_{\text{vN}}(A)$ is equal to the entanglement wedge cross-section $E_{W}$ in AdS spacetime, which is conjectured to be the holographic dual of the entanglement of purification. 
We show such constructed entanglement entropy has a phase transition.
The ordinary entanglement entropies of critical and non-critical QFTs are simply limits of the two phases.
\end{abstract}

\maketitle

\section*{Introduction}

Entanglement has fundamental importance for quantum information
and quantum gravity. In conformal field theories (CFTs), the entanglement
entropy classifies quantum entanglement between complementary parts
$A$ and $B$ in a pure entangled state $\psi_{AB}$, defined by the
von Neumann entropy 
\begin{equation}
S_{\text{vN}}(A)=-\mathrm{Tr}\rho_{A}\log\rho_{A}
\end{equation}
for a reduced density matrix $\rho_{A}=\mathrm{Tr}_{B}\vert\psi_{AB}\rangle\langle\psi_{AB}\vert$.
However, for a mixed states $\rho_{AB}\ne\vert\psi_{AB}\rangle\langle\psi_{AB}\vert$,
the entanglement entropy fails to characterize entanglement between
$A$ and $B$, as it yields the same values for entangled and unentangled
states. 

The typical resolution   involves purifying $AB$ by adding  auxiliary
systems $\bar{A}\bar{B}$, known as purification. The entanglement
of purification (EoP) \citep{Terhal:2002riz} is defined by minimizing
the entanglement entropy $S_{\text{vN}}(A\bar{A}:B\bar B)$ over all possible
purifications. 
Within the AdS/CFT correspondence \citep{Maldacena:1997re,Witten:1998qj,Gubser:1998bc},
the EoP is conjectured to be equal to the entanglement wedge cross-section
(EWCS) in the AdS bulk \citep{Takayanagi:2017knl}. 
However, the conjecture EoP $=$ EWCS is very hard to verify since in practice, 
carrying out the optimization over all possible purifications is almost not possible. Since the minimization of the EoP is very difficult, a canonical purification
\citep{Dutta:2019gen} is suggested as $\left|\sqrt{\rho_{AB}}\right\rangle \in\mathcal{H}_{A}\otimes\mathcal{H}_{\bar{A}}\otimes\mathcal{H}_{B}\otimes\mathcal{H}_{\bar{B}}$,
where $\bar{A}$ ( $\bar{B}$) is the reflection of $A$ ($B$).
The reflected entropy $S_{R}\left(A:B\right)$ is then defined as
the von Neumann entropy $S\left(\rho_{A\bar{A}}\right)$. However,
the reflected entropy still lacks a clear geometric interpretation,
since the bulk dual of $\left|\sqrt{\rho_{AB}}\right\rangle $ is
considered as the doubled entanglement wedge that remains elusive.
Another operationally defined and generally computable candidate is
the logarithmic negativity \citep{Calabrese:2012nk} whose explicit
form  is unknown \citep{Kudler-Flam:2018qjo},
since it depends on the conformal block that is computationally expensive.
Moreover, various measures
of mixed states are defined in different ways  \citep{Tamaoka:2018ned,Espindola:2018ozt,Harper:2019lff},
but they are all challenging to compute. 

In this letter, for a bipartite mixed state $\rho_{AB}$, 
in contrast to \emph{adding} extra auxiliary parts 
into the mixed state as the usual purifications do,
we show  by \emph{subtracting} the undetectable regions in
CFT$_2$, a pure entangled state $\psi_{AB}$
can be constructed on the doubly connected   plane (see
Fig. 1).
In this resulted pure state $\psi_{AB}$, it is natural to identify
the well defined
$S_{\text{vN}}(A)$ or $S_{\text{vN}}(B)$ as the entanglement measure
between $A$ and $B$. 
Our proposal is independent of purification
or four-point conformal block data, distinguishing it from  other
bipartite entanglement measures. We verify our proposal holographically
and find agreement with the EWCS in the AdS$_{3}$ bulk. 

\section*{Entanglement entropy in doubly connected region}

\begin{figure}[h]
\includegraphics[scale=0.38]{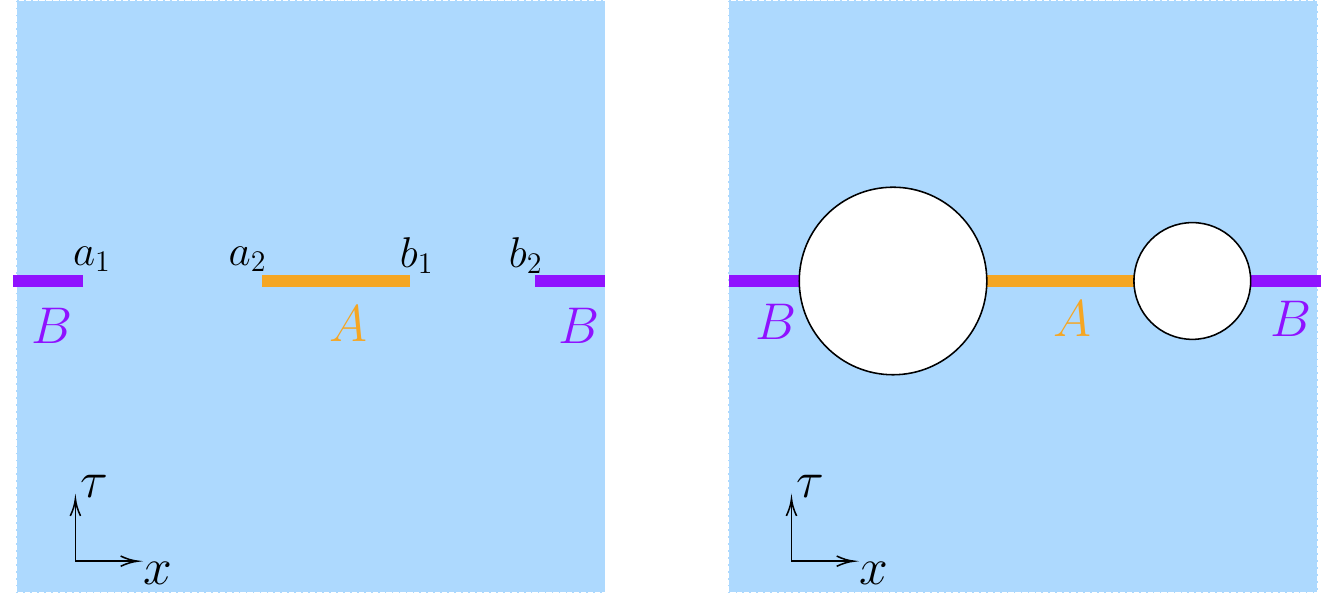}

\caption{Left panel: At $\tau=0$, two subsystems $A$ and $B$ on the $x$-axis
are in a mixed state $\rho_{AB}$. Right panel: By removing the undetectable parts with two discs having different radii, we
obtain a two-holed plane, in which two subsystems $A$ and $B$ are
now in a pure entangled state $\psi_{AB}$. The blue shaded region
represents the Euclidean path integral.\label{fig:UV-IR-regulators}}
\end{figure}

In a CFT$_2$, consider two disjoint subsystems $A=(a_{2},b_{1})$ and
$B=(-\infty,a_{1})\cup(b_{2},\infty)$ at time $\tau=0$, such that
$a_{1}<a_{2}<b_{1}<b_{2}$. The subsystems $A$ and $B$ are non-complementary
parts typically in a mixed state $\rho_{AB}$ (see the left panel
of Fig. \ref{fig:UV-IR-regulators}). 
In order to construct a pure state $\psi_{AB}$ from the mixed state $\rho_{AB}$,
we remove two discs in the Euclidean path integral region between $A$ and $B$,  obtaining
a two-holed plane, topologically a doubly connected region, 
as depicted in Fig. \ref{fig:UV-IR-regulators}. We then impose conformal invariant
boundary conditions on two edges of the two-holed plane, corresponding
to two boundary states $\vert a,b\rangle$ \citep{Cardy:1989ir}.
In this two-holed plane, two subsystems $A$ and $B$ are now complementary
parts in a pure entangled state $\psi_{AB}$, allowing us to define
the entanglement entropy $S_{\text{vN}}(A)$\footnote{The two holes can be equivalently understood as finite regulators}. 

\begin{figure}[t]
\includegraphics[scale=0.38]{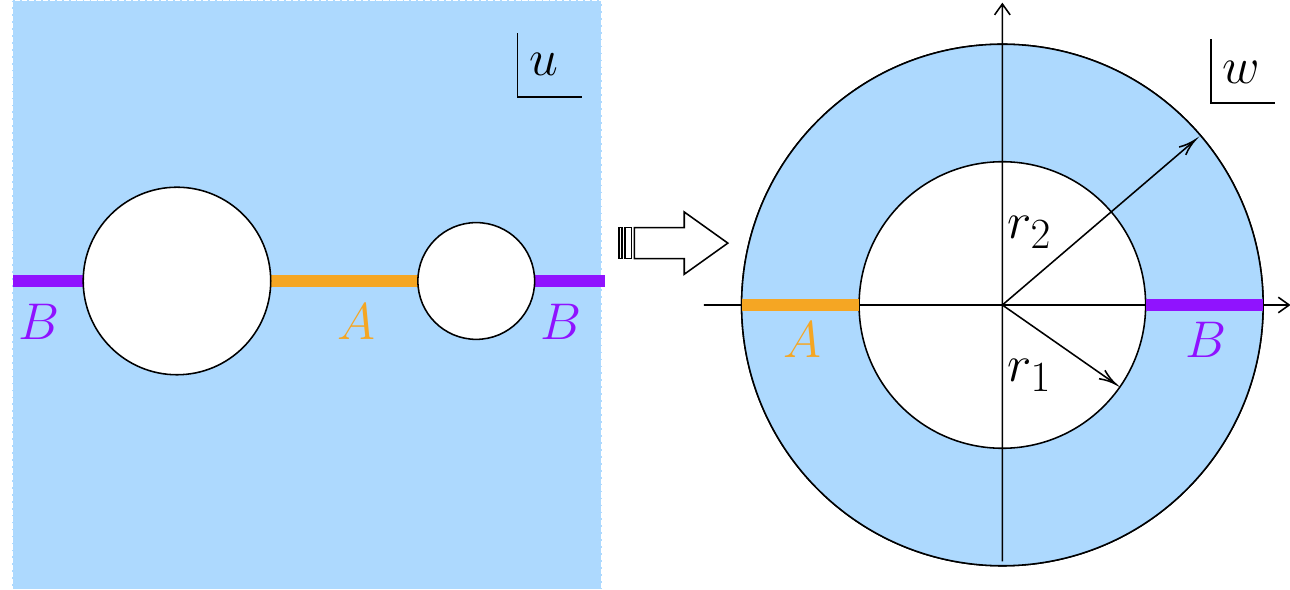}

\caption{The two-holed plane can be conformally mapped onto an annulus with
the inner radius $r_{1}$ and the outer radius $r_{2}$.\label{fig:annulus}\textcolor{blue}{{} }}
\end{figure}

Given that every doubly connected region can be conformally mapped
onto an annulus of the type illustrated in Fig. \ref{fig:annulus},
we first calculate the entanglement entropy $S_{\text{vN}}(A)$ between $A$ and $B$ in the annulus.
To this end, it is useful to define the width of the annulus,
\begin{equation}
W=\log\frac{r_{2}}{r_{1}},
\end{equation}
where $r_{1}$ and $r_{2}$ are the inner and outer radii, respectively.
The entanglement entropy $S_{\text{vN}}(A)$ in the annulus 
is defined through the R\'{e}nyi entropy:
\begin{eqnarray}
S^{(n)}(A) & = & \frac{1}{1-n}\log\mathrm{Tr}_{A}\rho_{A}^{n},\\
S_{\text{vN}}(A) & = & \lim_{n\rightarrow1}S^{(n)}(A),
\end{eqnarray}
where   $\mathrm{Tr}_{A}\rho_{A}^{n}$ is given by
\begin{equation}
\mathrm{Tr}_{A}\rho_{A}^{n}=\frac{Z_{n}}{Z_{1}^{n}}.
\end{equation}
Here, $Z_{1}$ is the partition function on a single cover of the
annulus, and $Z_{n}$ is the partition function on the $n$-sheeted
cover $\mathcal{M}_{n}$ obtained by sewing $n$ copies of $\mathcal{M}$
along $A$. Using the generator of scale transformations, the annulus
partition function can be expressed as \citep{Cardy:1989ir,Cardy:2004hm,Cardy:2016fqc}:
\begin{equation}
Z_{1}=e^{cW/12}\sum_{k}\langle a\vert k\rangle\langle k\vert b\rangle e^{-2\delta_{k}W},
\end{equation}
where $c$ is the central charge, $k$ denotes all allowed scalar
operators with dimensions $\delta_{k}$ inside the annulus, and $\vert a,b\rangle$
are boundary states. Similarly,
\begin{equation}
Z_{n}=e^{cW/12n}\sum_{k}\langle a\vert k\rangle\langle k\vert b\rangle e^{-2\delta_{k}W/n},
\end{equation}
since the replicated manifold $\mathcal{M}_{n}$ is conformally equivalent
to an annulus with the width $W_{n}=W/n.$ 
As usual, only the ground state ($k=0$) concerns us in calculating the entanglement entropy,
\begin{align}
Z_{1} & =e^{cW/12}\langle a\vert0\rangle\langle0\vert b\rangle,\label{eq:partition-1}\\
Z_{n} & =e^{cW/12n}\langle a\vert0\rangle\langle0\vert b\rangle,\label{eq:partition-n}
\end{align}
and $\mathrm{Tr}_{A}\rho_{A}^{n}$ read:
\begin{equation}
\mathrm{Tr}_{A}\rho_{A}^{n}=e^{\frac{c}{12}(\frac{1}{n}-n)W}\left(\langle a\vert0\rangle\langle0\vert b\rangle\right)^{1-n}.\label{eq:moments}
\end{equation}
Substituting this into the R\'{e}nyi entropy expression, we obtain:
\[
S^{(n)}(A)=\frac{c}{12}\left(1+\frac{1}{n}\right)W+g_{a}+g_{b},
\]
where $g_{a,b}=\log\langle a,b\vert0\rangle$ are the Affleck-Ludwig
boundary entropies \citep{Affleck:1991tk}, which  encode the information 
of the undetectable regions and are irrelevant. Thus the entanglement 
between subsystems $A$ and $B$ is quantified by the universal term as
\begin{equation}
S_{\text{vN}}(A:B)=\lim_{n\rightarrow1}S^{(n)}(A)=\frac{c}{6}W.
\label{eq:annulus W}
\end{equation}

Now, we can map the $u$-plane with two holes onto the annulus
in the $w$-plane via the conformal transformations:

\begin{equation}
v(u)=2\frac{u-a_{1}}{a_{2}-a_{1}}-1,\quad w(v)=\frac{v-\gamma}{\gamma v-1},
\end{equation}
with
\begin{gather}
\gamma=\frac{1+\alpha\beta+\sqrt{(1-\alpha^{2})(1-\beta^{2})}}{\alpha+\beta},\nonumber \\
\alpha=v(b_{1}),\quad\beta=v(b_{2}).
\end{gather}
The width of the annulus is
\begin{equation}
W=\log\frac{\alpha-\beta}{1-\alpha\beta+\sqrt{(1-\alpha^{2})(1-\beta^{2})}}.
\label{eq:AB W}
\end{equation}
In terms of the cross ratio 
\begin{equation}
z=\frac{(a_{2}-a_{1})(b_{2}-b_{1})}{(b_{1}-a_{2})(b_{2}-a_{1})},
\label{eq:cross ratio}
\end{equation}
plugging (\ref{eq:AB W}) into (\ref{eq:annulus W}), we get
\begin{equation}
S_{\text{vN}}(A:B)=\frac{c}{6}\log\left[1+\frac{2}{z}+2\sqrt{\frac{1}{z}\left(\frac{1}{z}+1\right)}\right].\label{eq:CD-RG}
\end{equation}
In the limit   $b_{1}-a_{2}=\ell$ and $a_{2}-a_{1}=b_{2}-b_{1}=\epsilon \to 0 $ as a UV regulator,
this entanglement entropy simplifies to the famous single interval result
\begin{equation}
\ensuremath{S_{\text{vN}}(A:B)=\frac{c}{3}\log\frac{\ell}{\epsilon}}.
\end{equation}
%

\begin{figure}[t]
\includegraphics[scale=0.38]{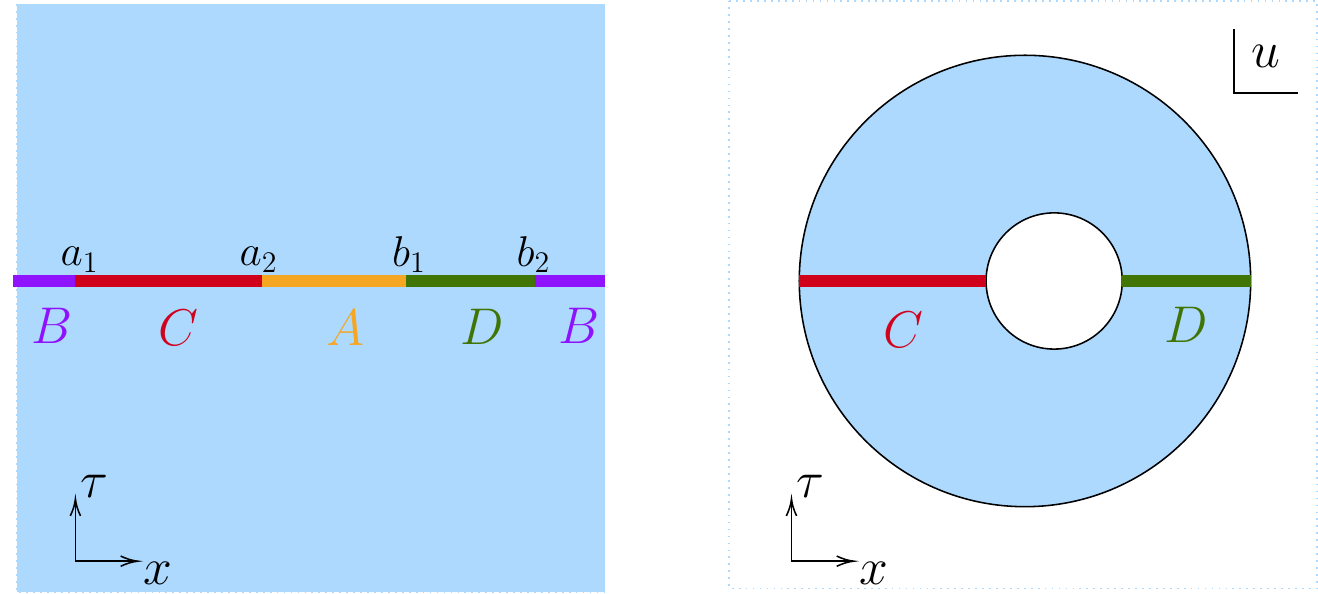}

\caption{Left panel: Four subsystems $ABCD$ in the $u$-plane. Right panel:
by subtracting the undetectable parts $A$ and $B$ with two discs, we obtain an
asymmetric annular region in which  $C$ and $D$ are
in a pure entangled state $\psi_{CD}$.\label{fig:Asymmetric-RG}}
\end{figure}

On the other hand, for the same system, it is illuminating to switch the roles of $A$, $B$ and 
those of the undetectable regions $C$, $D$, as depicted in Fig. \ref{fig:Asymmetric-RG}. 
So, we are now concerned with the entanglement between
$C\in(a_{1},a_{2})$ and $D=(b_{1},b_{2})$, while $A$ and $B$ are undetectable regions. 
Following the similar procedure, we subtract $A$ and $B$ with two discs. 
The resulted manifold is an asymmetric annulus, where $C$ and $D$ become 
complementary parts in a pure entangled state $\psi_{CD}$.


We can map the asymmetric annular region in the $u$-plane onto an
annulus in the $w$-plane using the following conformal transformations:

\begin{equation}
\tilde{v}(u)=2\frac{u-a_{1}}{b_{2}-a_{1}}-1,\quad w(\tilde{v})=\frac{\tilde{v}-\tilde{\gamma}}{\tilde{\gamma}\tilde{v}-1},
\end{equation}
with
\begin{gather}
\tilde{\gamma}=\frac{1+\tilde{\alpha}\tilde{\beta}+\sqrt{(1-\tilde{\alpha}^{2})(1-\tilde{\beta}^{2})}}{\tilde{\alpha}+\tilde{\beta}},\\
\tilde{\alpha}=\tilde{v}(b_{1}),\quad\tilde{\beta}=\tilde{v}(a_{2}).
\end{gather}
The width of the annulus is
\begin{equation}
\tilde{W}=\log\frac{1-\tilde{\alpha}\tilde{\beta}+\sqrt{(1-\tilde{\alpha}^{2})(1-\tilde{\beta}^{2})}}{\tilde{\alpha}-\tilde{\beta}}.
\end{equation}
In terms of the cross ratio (\ref{eq:cross ratio}),
the entanglement entropy between $C$ and $D$ is 
\begin{eqnarray}
S_{\text{vN}}(C:D)&=&\frac{c}{6}\tilde{W}\nonumber\\
&=&\frac{c}{6}\log\left[1+2z+2\sqrt{z\left(z+1\right)}\right].\label{eq:Asymmetric-EE}
\end{eqnarray}
In the limit   $b_{1}=-a_{2}=\epsilon\rightarrow0$ and $b_{2}=-a_{1}=\xi\rightarrow\infty$,
we reproduce the half-space entanglement entropy 
\begin{equation}
S_{\text{vN}}(C:D)=\frac{c}{6}\log\frac{\xi}{\epsilon}.
\end{equation}

\section*{Holographic Interpretation}

\begin{figure}[t]
\includegraphics[scale=0.6]{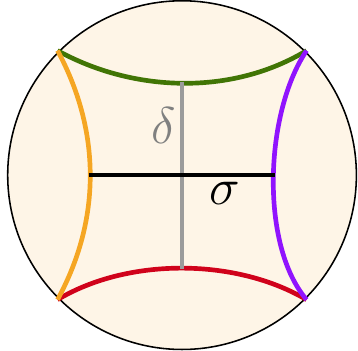}

\caption{In the hyperbolic disk, a quadrilateral is bounded by colored geodesics,
and the only two perpendicular geodesics (black and gray lines) satisfy the
equation $\sinh\frac{\delta}{2}\sinh\frac{\sigma}{2}=1$, where $\sigma$
and $\delta$ denote their lengths.\label{fig:hyperbolic-disk}}
\end{figure}

Notably, $S_{\text{vN}}(C:D)$ is related to $S_{\text{vN}}(A:B)$
by
\begin{equation}
z\rightarrow\frac{1}{z},
\end{equation}
and satisfy
\begin{equation}
\sinh\left[\frac{3}{c}S_{\text{vN}}(A:B)\right]\sinh\left[\frac{3}{c}S_{\text{vN}}(C:D)\right]=1.\label{eq:hyper condition}
\end{equation}
This equation is an exclusive constraint in hyperbolic geometry \citep{buser2010}
for the only two perpendicular geodesics inside a quadrilateral, as
depicted in Fig. \ref{fig:hyperbolic-disk}. This observation is reminiscent
of the AdS/CFT correspondence, where it has been conjectured that
the minimal cross-section of the entanglement wedge $E_{W}$ in the
AdS bulk is a holographic dual of the EoP in a given bipartite mixed
state \citep{Takayanagi:2017knl}. With the same setup, the EWCS could
be computed in the AdS$_{3}$ bulk, and we immediately find that the
EWCS precisely agrees with the entanglement entropy $S_{\text{vN}}$
in the doubly connected region:
\begin{eqnarray}
E_{W}(A:B) & = & S_{\text{vN}}(A:B),\label{eq:asymmetric-EWCS}\\
E_{W}(C:D) & = & S_{\text{vN}}(C:D).
\end{eqnarray}
Thus, the entanglement entropy $S_{\text{vN}}$ in the doubly connected
region should be regarded as an alternative to   EoP.

In addition, our proposal provides a novel interpretation of the entanglement
phase transition for mixed states. From Eq. (\ref{eq:hyper condition}),
we identify a critical point:
\begin{equation}
S_{\text{vN}}(A:B)=S_{\text{vN}}(C:D)=S_{\text{vN}}^{*},
\end{equation}
where the critical value
\begin{equation}
S_{\text{vN}}^{*}=\frac{c}{3}\mathrm{arcsinh}\,1=\frac{c}{6}\log\left(3+2\sqrt{2}\right)\label{eq:critical-value}
\end{equation}
matches the phase transition point of the EWCS \citep{Takayanagi:2017knl}.
This entanglement phase transition is usually interpreted as
the mutual information $I(A:B)=\frac{c}{6}\log z$ between two subsystems,
vanishing when $z\le1$. We now see from our results this is precisely a transition of the dominant
contribution between $s$-channel and $t$-channel \citep{Hartman:2013mia}.
It indicates that the entanglement transition near $z=1$ signifies
a transition from entanglement between $A$ and $B$ to entanglement
between $C$ and $D$, or vice versa.

\section*{CONCLUSION AND DISCUSSION}

In this letter, we proposed an alternative to purification for bipartite
mixed states in CFT$_2$. Our proposal is completely determined by
the definition of von Neumann entropy and the replica trick, independent
of any purification or four-point conformal block data. 
We confirmed the proposed $S_{\text{vN}}$ is the dual of EWCS and
equal to  EoP.

Our choice of regulators, by removing discs, is \emph{canonical},
although other choices with different shapes may exist. The width
$W$ is a \emph{conformal invariant}; if two annuli have different
widths, they cannot be conformally mapped onto each other. The set
of all doubly connected regions falls into classes of conformally
equivalent regions, with each class characterized by the width of
that class. Therefore, any other possible regulators with different
shapes must be conformally equivalent to our choice.

Our proposal suggests that the von Neumann entropy in
the multi-connected regions may correspond to an entanglement measure
for multipartite mixed states. Having successfully extracted the EWCS
from the annulus CFT, it is of interest to further study the connections
between the entanglement wedge and the CFT in  multi-connected regions.

In \citep{Jokela:2019ebz}, the EWCS for two parallel strips with equal widths
$l$ in AdS$_{d+1}$ are given. It turns out our proposal can be used to find 
the dual entanglement entropy in CFT$_4$.

\vspace*{3.0ex}
\begin{acknowledgments}
\paragraph*{Acknowledgments.} 
We are very grateful to Song He, Bo Ning, Ronak M Soni, Tadashi Takayanagi, Qiang Wen, Runqiu Yang and Yang Zhou for reading the first version of this work. Their comments or suggestions help us to improve the manuscript substantially. 
This work is supported in part by NSFC (Grant No. 12105191, 12275183 and 12275184).
\end{acknowledgments}

\bibliographystyle{unsrturl}
\bibliography{Alternative_to_Purification_0613_PRL_V1_Jiang}

\end{document}